# Unraveling the Excitonic Transition and Associated Dynamics in Confined Long Linear Carbon-chains with Time-Resolved Resonance Raman Scattering

*Jingyi Zhu\*, Robin Bernhardt, Weili Cui, Raphael German, Julian Wagner, Boris V. Senkovskiy, Alexander Grüneis, Thomas Pichler, Rulin Liu, Xi Zhu, Paul H. M. Van Loosdrecht\* and Lei Shi\**

Dr. J. Zhu, R. Bernhardt, R. German, J. Wagner, Dr. B. V. Senkovskiy, Prof. A. Grüneis, Prof. P. H. M. Van Loosdrecht

University of Cologne, Physics institute 2, 50937, Germany

E-mail: jzhu@ph2.uni-koeln.de ; pvl@ph2.uni-koeln.de

Prof. L. Shi

School of Materials Science and Engineering, Sun Yat-sen University, 510275, China
Faculty of Physics, University of Vienna, Vienna 1090, Austria
E-mail: shilei26@mail.sysu.edu.cn

W. Cui, Prof. Thomas Pichler

Faculty of Physics, University of Vienna, Vienna 1090, Austria

R. Liu, X. Zhu

The Chinese University of Hong Kong (Shenzhen), Shenzhen, Guangdong 518172, China

**Keywords: long linear carbon-chains, carbon nanotube, excitonic population, energy transfer, time-resolved Raman**

**Abstract:** Long linear carbon-chains have been attracting intense interest arising from the remarkable properties predicted and their potential applications in future nanotechnology. Here we comprehensively interrogate the excitonic transitions and the associated relaxation dynamics of nanotube confined long linear carbon-chains by using steady state and time-resolved Raman spectroscopies. The exciton relaxation dynamics on the confined carbon-chains occurs on a hundreds of picoseconds timescale, in strong contrast to the host dynamics that occurs on a few picosecond timescale. A prominent time-resolved Raman response is observed over a broad energy range extending from 1.2 to 2.8 eV, which includes the strong Raman resonance region around 2.2 eV. Evidence for a strong coupling between the chain and the nanotube host is found from the dynamics at high excitation energies which provides a clear evidence for an efficient energy transfer from the host carbon nanotube to the chain. Our experimental study presents the first unique characterization of the long linear carbon-chain exciton dynamics, providing indispensable knowledge for the understanding of the interactions between different carbon allotropes.

## 1. Introduction

Long linear carbon-chain (LLCC), with its infinite long form termed as carbyne,[1] is a one-dimensional $sp^1$-hybridized carbon allotrope unique from $sp^2$-hybridized graphene and $sp^3$-hybridized diamond. The backbone of carbyne is constructed by either (−C≡C−) or (=C=) as a repeating base element, and the unique *sp*-hybridization endows it a truly straight one-dimensional crystalline material. Although carbyne has been proposed to exist long time ago,[2] experimental synthesis and identification of LLCCs was only realized in recent years. Nevertheless, many superior properties of carbyne have been predicted by theoretical calculations, such as a hardness higher than diamond,[3] high thermal and electronic conductivities,[4] as well as remarkable high bending and strain tolerances.[5] These intriguing theoretical predictions have led to an intensive effort over the past few decades to synthesize LLCCs.[6] Due to the high reactivity of the *sp*-hybridized bonds, synthesis of long and stable chains is still challenging. Without special protection and isolation, carbyne easily decays by forming either *sp²* or *sp³*-carbon allotropes through crosslinking between adjacent chains or through cycloaddition reactions.[7] Different strategies have been adopted to stabilize LLCCs, the main ones being applying end-capping groups[6f] or confining in host nanotubes.[6e] Recently, LLCCs incorporating more than 6000 carbon atoms[8] have been synthesized in the double-walled carbon nanotube (DWCNT) hosts. For these confined chains, it has been shown that their properties depend strongly on the interaction with their hosts, and only weakly on the length of the chain.[9]

Despite the progress in synthesis, the experimental characterization of LLCCs is still limited. This is in particular true for optically excited non-equilibrium states. The presence of other carbon allotropes[10] makes identifying the LLCC properties a challenging task. To date, the most successful techniques to address the properties are transmission electron microscopy and Raman scattering spectroscopy.[8, 11] In particular Raman spectroscopy has proven to be a convenient tool to study the vibrational properties of LLCCs.[12] Taking it one step further, time-resolved Raman spectroscopy (TRRS) has demonstrated to be a powerful method to investigate both the vibrational relaxation and the electronic population dynamics for carbon based nanomaterials.[13] Especially, performing TRRS under strong resonance conditions allows addressing the electronic excitation dynamics directly from the differential Stokes signals.[13e] Here, we report on the temperature and wavelength dependent optical excitation relaxation dynamics in the confined LLCCs and the excited state coupling to the hosts by monitoring the optical phonons using TRRS under resonant conditions. Our findings are pivotal to understand the fundamental electronic and optical properties of LLCCs towards potential applications. To the best of our knowledge, this is the first experimental report on non-equilibrium dynamics for a combined system of the $sp^1/sp^2$-hybridized carbon allotropes.

## 2. Results and discussion

Figure 1(a) depicts a LLCC encapsulated in a host DWCNT (LLCC@DWCNT) and an empty DWCNT. Due to the Peierls distortion,[14] the LLCC has a stable polyyne-like structure $(-C≡C-)_n$ as opposed to a cumulene-like structure $(=C=)_n$.[15] The unit cell of the polyyne-like structure contains two carbon atoms providing two electrons, which completely fill up the valence band. This results in a semiconducting nature of the LLCCs with excitons as the fundamental optical excitations. Raman scattering spectra of the samples (see figure 1(c)) show the presence of the typical G-modes at ~1590 cm$^{-1}$ for both LLCCs@DWCNTs as well as pristine DWCNTs. The weak mode observed for both samples at ~1750 cm$^{-1}$ most likely originates from combination scattering involving the radial breathing mode (RBM, spectra of the RBM modes can be found in Supporting information Figure S1) and the G-mode, similar to that observed in graphite and other nanotubes.[16] The presence of the linear carbon-chains is evidenced by the appearance of Raman peaks at ~1840 cm$^{-1}$ for LLCCs@DWCNTs corresponding to the typical stretching vibrations of LLCCs.[6e, 8, 11b, 17] This carbon-chain band is well separated from those of DWCNTs, providing a unique experimental access to investigate the excited state dynamics in the time-resolved resonant measurements. Before performing the time-resolved Raman measurements, we first characterize the steady state resonant Raman response of LLCCs@DWCNTs over a wide range of photon energies. Detailed Raman spectra recorded at various wavelengths are presented in Figure S2 in the supporting information. The resonance profile obtained from the phonon peak centered around 1840 cm$^{-1}$ is presented in Figure 1(d). It shows a clear resonance around 2.2 eV with a FWHM (Full width at half maximum) of 0.3 eV, and a tail extending below 1.8 eV, consistent with previously reported results of confined LLCCs.[17a, 18]

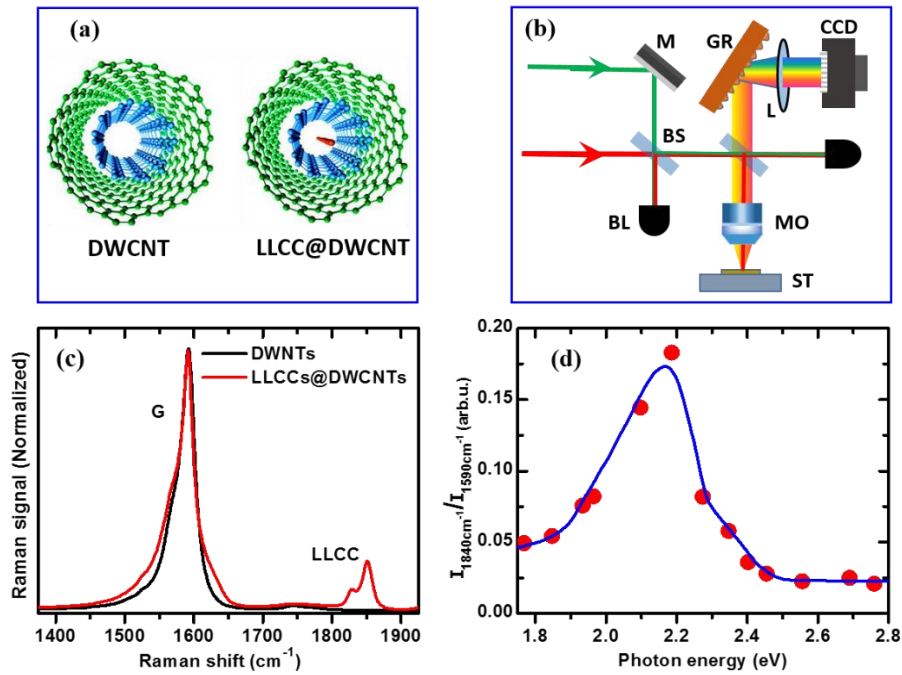

**Figure 1.** (a) Illustration of a DWCNT and a confined LLCC@DWCNT. (b) Sketch of the time-resolved Raman scattering setup. Green, red, and broad colorful lines indicate the Raman probe, the pump, and the Raman scattered signal, respectively. M: mirror, GR: grating, CCD: charge coupled device, BS: beam splitter, BL: beam block, MO: microscope objective, ST: sample stage. (c) Steady-state Raman spectra of DWCNTs and LLCCs@DWCNTs measured using a 532 nm continuous laser as excitation source. (d) Resonance Raman profile of LLCCs obtained from the intensity of mode 1840 cm$^{-1}$ relative to 1590 cm$^{-1}$, the blue solid line is a guide to the eyes.

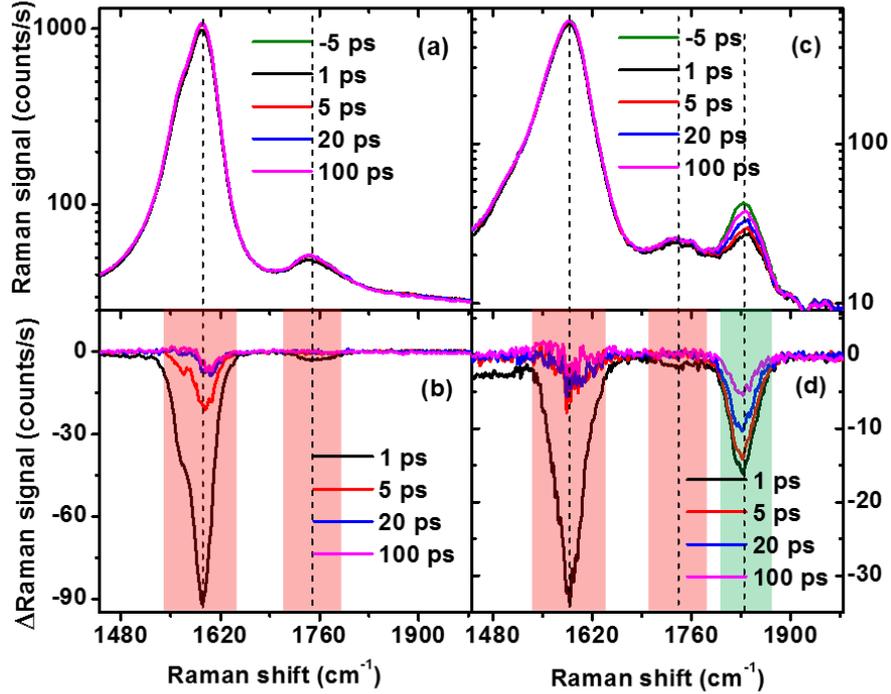

**Figure 2.** Time-resolved Raman scattering spectra of the pristine DWCNTs and LLCCs@DWCNTs recorded at various pump-probe delay times. (a) Raman spectra of pristine DWCNTs. (b) Difference Raman spectra of pristine DWCNTs. (c) Raman spectra of LLCCs@DWCNTs. (d) Difference Raman spectra of LLCCs@DWCNTs. To highlight the relatively small amplitude from the LLCC mode, the spectra in (a) and (c) are plotted using a logarithmic intensity scale.

In the time-resolved Raman measurements, we first applied a resonant pump at 2.1 eV and a probe at 2.4 eV (512 nm). The transient changes in the Raman response have been recorded in the 1500-2000 cm$^{-1}$ spectral window covering both the DWCNT G-mode and the LLCC stretching mode. The room temperature time-resolved Raman spectra of pristine DWCNTs and LLCCs@DWCNTs are recorded using a moderate pump intensity of around 25 μJ cm$^{-2}$, and are presented in Figure 2(a) and (c), respectively. Immediately after optical excitation, the intensity of all phonon modes show a significant reduction for both samples. This reduction is a direct consequence of bleaching of the ground state by the pump excitation pulse, which reduces the resonant enhancement of the Raman probing signals, thus providing a direct measurement of electronic excited state populations.[13e, 19] To demonstrate the pump induced effect more clearly, the lower panels of Figure 2b and d show the difference Raman spectra for DWCNTs and LLCCs@DWCNTs obtained by subtracting the -5 ps spectra. The DWCNT G-bands around 1590 cm$^{-1}$ and 1750 cm$^{-1}$ show a rapid recovery within a few ps for both pristine DWCNTs and LLCCs@DWCNTs. The initial intensity reduction ratio of the LLCC stretching mode (1840 cm$^{-1}$) of LLCCs@DWCNTs is significantly larger than that of the DWCNT modes, indicating a stronger suppression of the resonance enhancement for LLCC. The most surprising observation, however, is that the LLCC mode recovers much slower than the DWCNT modes, taking about 100 ps, suggesting a relaxation mechanism different from the DWCNT modes.

Figure 3 shows the time evolution of the various phonon modes obtained by integrating the transient signal over the corresponding spectral region (shaded regions in Figure 2b and d). The 1590 and 1750 cm$^{-1}$ modes show identical multi-exponential recovery dynamics for which a global fitting yields that 80% of the signal decays with a time constant of 0.3 ps, and 20% with a time constant of 2.7 ps. Similar relaxation dynamics were observed for the 1590 cm$^{-1}$ peak for carbon nanotubes made in different batches (Supporting information Figure S3). We note that the observed fast decay dynamics for the DWCNTs are consistent with previous observations on multi-walled carbon nanotubes which reported a similarly short lifetime of around 1 ps.[20] It is clear that the decay dynamics of the LLCC stretch mode is substantially slower than that of the DWCNT modes. Fitting a multi-exponential decay function to the data yields time constants of ~7 and ~80 ps, with amplitudes of 35% and 55% respectively. A small part of the transient signal (10%) shows no decay at all in the measured time window. Alternatively, an exponential decay with a log-normal distribution of the relaxation rate can also fit the observed dynamics perfectly (Supporting information Figure S4). More discussion on the mechanism of the relaxation dynamics and the multi-exponential nature will follow later on when presenting the temperature and excitation intensity dependent measurements. For convenience in describing the dynamics, we adhere to the multi-exponential fitting procedure for all decay dynamics discussed in the following.

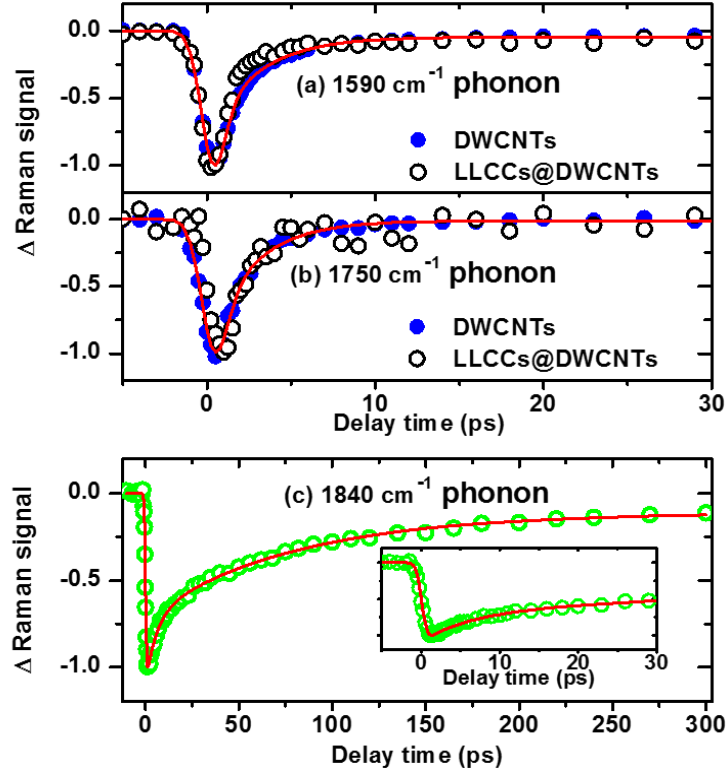

**Figure 3.** Normalized relaxation dynamics of phonon Raman scattering in pristine DWCNTs and LLCCs@DWCNTs for the phonon bands at (a) 1590 cm$^{-1}$, (b) 1750 cm$^{-1}$, and (c) 1840 cm$^{-1}$. The inset in (c) shows an expanded view of the short time range. Symbols are experimental data and red lines are multi-exponential fits.

Since we always excite the hosts and the confined LLCCs simultaneously, the much slower dynamics of the LLCC mode compared to their hosts suggest that excitation energy should flow from DWCNTs down to LLCCs if there is energy transfer between them. Due to the ps time resolution of our Raman probing, the current single excitation energy data does not allow to resolve a direct energy transfer between the DWCNTs and the LLCCs if the process is very fast. To further investigate the intricate optical transition and relaxation dynamics, we carried out excitation energy dependent measurements of the LLCCs relaxation dynamics from the 1840 cm$^{-1}$ mode. While fixing the probing energy at 2.4 eV, the excitation photon energy was tuned over a wide range from 2.8 to 1.2 eV. A few representative relaxation dynamic curves are presented in Figure 4a, in which all curves are normalized to the pre-zero Raman signal and to the pump excitation photon density. The maximum reduced signal amplitude (around 1 ps) is converted into the ground state bleaching ratio by taking the changes of Raman response into account as described in previous reports[13e, 21], and is presented in Figure 4b. More decay curves at specific excitation energies are presented in Figure S5 in the Supporting information.

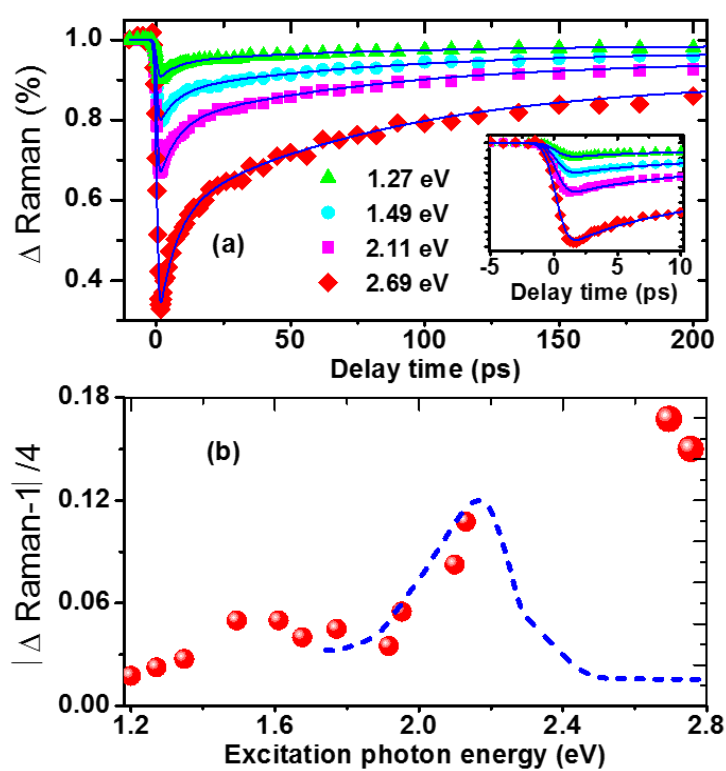

**Figure 4.** Excitation energy dependence of the relaxation dynamics of the LLCC mode and the population efficiency. (a) Representative decay dynamics observed for excitation energies far below (1.27 eV, 1.49 eV), at (2.11 eV), and far above (2.69 eV) the resonant transition energy. The inset shows the short time dynamics. (b) Extracted optical bleaching percentage of the ground state at 1 ps. The blue dashed line is the resonance profile taken from Figure 1(d).

Several significant points are revealed from the photon energy dependent excitation results. First, dynamics at all the different excitation energies are very similar and can be globally described well using three decay components with the same time constants, with small variations in the component amplitudes (Supporting information Figure S5 and Figure S6). This indicates that, independent of the excitation energy, all signals originate from excitation of the same species. Second, in the middle energy range, the resonance profile obtained from the time resolved experiments fit well to that from the steady state measurements, confirming the strong resonance effect around 2.2 eV. Third, at lower excitation energies, we still observe a substantial LLCC response signal. Understanding this phenomenon is very challenging due to absence of data on the excited state spectra and dynamics of the confined LLCCs from any other experiments. The much broader responses may come from some low energy dark states of the LLCCs that becomes bright when interacting with the DWCNTs, from some intrinsic resonant states with lower lower density of the LLCCs below 2.2 eV, or from a very fast energy transfer from the hosts to the LLCCs. High time-resolution experiments are required to further distinguish these possibilities. Fourth and most significantly, on the high energy side around 2.7 eV, where the steady state measurements do not show an observable resonance, the observed population effect is even more enhanced. This unambiguously indicates that there is significant energy transfer from the excited host DWCNTs to the confined LLCCs. Since the photoluminescence states of the host DWCNTs lie in in the infrared region [22] one does not expect a Förster-type energy transfer but rather a Dexter-type. A Dexter-type energy transfer is indeed expected in view of the strong overlapping of the molecular orbitals due to the small radius of the inner tube of the DWCNTs. Intriguingly, the population amplitude at ~2.7 eV excitation is even larger than that at the resonance excitation of LLCCs at ~2.1 eV. A possible origin for this surprising observation lies in the much higher density of states available for both the LLCCs (dark state) and the host tubes in the near UV region, making the near UV excitation very efficient in bleaching the LLCC through DWCNT-LLCC energy transfer which subsequently decays to the lowest excited LLCC state at 2.2 eV thereby bleaching the resonance enhancement of the LLCC phonon response.

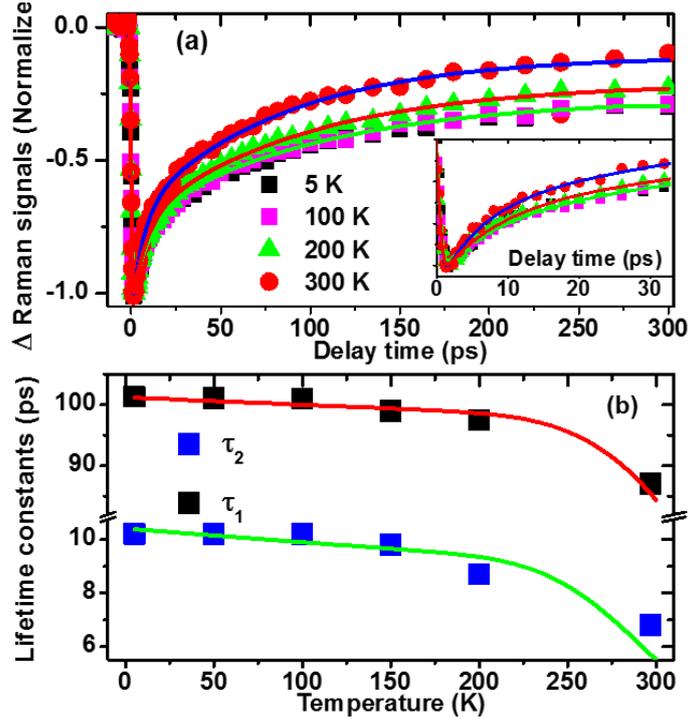

**Figure 5.** Temperature dependent excited state dynamics of the LLCC mode. (a) Normalized decay dynamics at various temperatures (symbols: data, lines: multi-exponential fits). The inset shows the short time dynamics. (b) Temperature dependence of the extracted time constants (dots), lines are fits using a statistical phonon occupation model (see text).

We now return to recovery mechanisms of the observed exciton relaxation dynamics of LLCCs. To shed some more light on this we have performed temperature dependent measurements on the relaxation dynamics at the fixed excitation energy of 2.1 eV and at around of ~25 μJ cm$^{-2}$. As shown in Figure 5, the dynamics is nearly temperature independent below 150 K but speeds up only at for higher temperatures. Figure 5(b) shows the temperature dependence of the two time constants obtained from multi-exponential fits. Since no luminescence has been detected at any temperature, we can assume that the dominant decay mechanism is non-radiative and has most likely, in view of the temperature dependence, an optical phonon-assisted nature. We therefore fitted the temperature dependent decay rates using a phonon occupation model:

$$\tau^{-1}(T) = \Gamma_0 + \Gamma_1(T) + \Gamma_2(T) \qquad (1)$$

where $\Gamma_0$ represents a constant contribution originating from the impurities and defects, $\Gamma_1$ (T) represents a contribution from low energy acoustical phonons, which is proportional to T, and $\Gamma_2$ (T) represents the contribution from high frequency optical phonons, which is proportional to the Bose-factor $1/[\exp(\hbar\omega_0/k_BT) -1]$. From the fit we obtain $\omega_0 \approx 1800$ cm$^{-1}$ for $\tau_1$ and $\omega_0 \approx 1820$ cm$^{-1}$ for $\tau_2$, indicating that the high energy optical phonon modes of the LLCCs play an extra role in the exciton relaxation dynamics at higher temperatures.

For the non-exponential nature of the recovery dynamics, there could be several reasons: 1) diffusion assisted recombination at defects or sample ends as those observed in other one-dimensional systems [23]; 2) exciton-exciton annihilation reaction; and 3) a wide range distribution of the LLCCs with different length and of the different type host DWCNTs. [22, 24] In the first two situations the dynamics is strongly dependent on the exciton diffusion. However, the observation that the dynamics is nearly temperature independent below 150 K rule out this possibility, since the diffusion constant is expected to be linearly dependent on temperature according to the Einstein relation. We also performed a set of experiments varying the pump power-density by one order of magnitude. These experiments show identical dynamics (Supplementary Figure S7), confirming once again the lack of exciton diffusion as well as the lack of an important contribution from exciton-exciton annihilation processes to the dynamics. Considering that the dynamics could also be well described by an exponential decay with a log-normal distribution of the relaxation rate (Supplementary Figure S4), we thus assign the origin of the non-exponential dynamics to inhomogeneity which may include variations of the length of LLCCs, different type of host DWCNTs, as well as possible variations in bonding between the host and the LLCC.

## 3. Conclusion

In conclusion, we have systematically investigated the optical transition and the excitonic population dynamics in the LLCCs@DWCNTs by steady state and time-resolved resonance Raman spectroscopy. Combining the advantages of the good resolution capability in both frequency and time-domain, we have successfully discriminated dynamic signals from different constituents and observed short and long-lived excited state population lifetime of the host DWCNTs and the confined LLCCs, respectively. The optical response of the LLCCs was observed to be in a wide energy range far below 2.2 eV, and an efficient Dexter-type energy transfer was identified from the excited DWCNTs to the LLCCs with excitation energy at the near UV side (~2.7 eV) by comparing to the steady state resonant transition of LLCCs, indicating the presence of strong excited state interactions between the two carbon allotropes. The observed temperature dependence of the dynamics show that the exciton recombination is dominated by defects and optical phonon-assisted processes, whereas the power dependent experiments showed that exciton-exciton recombination does not play any significant role.

## 4. Experimental Methods

*Sample preparation*: Samples consisting of carbon-chains inside the double wall carbon nanotubes (LLCCs@DWCNTs) have been produced by annealing the enhanced direct injection pyrolytic synthesized single-walled carbon nanotubes (eDIPS SWCNTs) at 1460 °C in high vacuum (<$10^{-6}$ mbar) for 1 hour. Details of the synthesis and characterization can be found in references. [9b, 25]

*Steady state Raman scattering spectroscopy:* Typical steady state Raman spectra of the LLCCs@DWCNTs and the DWCNTs were measured using a commercial Raman spectrometer (Princeton Instruments IsoPlane) equipped with a deep-cooled back-illuminated CCD. A continuum wave 532 nm laser with 0.5 mW power was used for experiments. A microscope objective with numerical aperture 0.4 (Olympus MPLFLN, 20x) was used to focus the laser beam onto the sample. Wavelength dependent resonance Raman measurements were performed by using the time-resolved setup described below.

*Time-resolved Raman scattering spectroscopy:* Details of the time-resolved resonance Raman scattering spectroscopy method have been described elsewhere.[21, 26] Briefly, an integrated ultrafast laser system (Light Conversion PHAROS) with two outputs of the fundamental 1030 nm pulses (compressed 300 fs and chirped 150 ps) pumps two optical parametric amplifiers (Light Conversion) to generate wavelength tunable laser pulses for selective excitation (~300 fs) and narrow-bandwidth laser pulses for Raman probing (~1.5 ps), respectively. The widely tunable pump pulse allows to cover a photon energy range from the near ultraviolet to the near infrared. This output is also used as light source for the steady state resonance Raman measurement mentioned above. The Raman probe laser pulses were sent to a 4*f*-pulse shaper to narrow the line width down to ~10 cm$^{-1}$ and suppress background light. The major scheme of the time-resolved Raman scattering spectroscopy setup/experiment is illustrated in figure 1(b). To reduce background signal originating from the strong optical pump, a perpendicular linear polarized pump and probe laser pulses are used in the experiments with the pump light blocked by a polarizer in front of the spectrometer. In order to avoid possible optical damage of the sample and to improve the signal to noise ratio, the laser spot size on the sample was defocused to ~80 μm in diameter. For all measurements, samples were mounted in a Helium cooled cold-finger cryostat with vacuum condition maintained at ~10$^{-6}$ mbar.

**Acknowledgements**

The work is supported by the Deutsche Forschungsgemeinschaft (DFG, German Research Foundation) via the project No. 277146847 - CRC1238: Control and Dynamics of Quantum Materials. L. S. acknowledges the financial support from the National Natural Science Foundation of China (51902353) and the Natural Science Foundation of Guangdong Province (No. 2019A1515011227). X.Z. thanks the National Natural Science Foundation of China (Grants 22075240).

**References**

[1]     a) F. Diederich, Y. Rubin, *Angew Chem Int Edit* **1992**, 31, 1101; b) F. Diederich, *Nature* **1994**, 369, 199; c) F. P. Bundy, W. A. Bassett, M. S. Weathers, R. J. Hemley, H. K. Mao, A. F. Goncharov, *Carbon* **1996**, 34, 141; d) L. Kavan, *Chem Rev* **1997**, 97, 3061.

[2]     A. von Bayer, *Berichte der deutschen chemischen Gesellschaft* **1882**, 15, 50.

[3]     L. Itzhaki, E. Altus, H. Basch, S. Hoz, *Angewandte Chemie-International Edition* **2005**, 44, 7432.

[4]     a) M. C. Wang, S. C. Lin, *Sci Rep-Uk* **2015**, 5; b) Y. Zhu, H. C. Bai, Y. H. Huang, *Chemistryopen* **2016**, 5, 78.

[5]     M. J. Liu, V. I. Artyukhov, H. Lee, F. B. Xu, B. I. Yakobson, *Acs Nano* **2013**, 7, 10075.

[6]     a) F. Cataldo, *Polym Int* **1999**, 48, 15; b) R. J. Lagow, J. J. Kampa, H. C. Wei, S. L. Battle, J. W. Genge, D. A. Laude, C. J. Harper, R. Bau, R. C. Stevens, J. F. Haw, E. Munson, *Science* **1995**, 267, 362; c) F. Cataldo, *Carbon* **2004**, 42, 129; d) K. H. Xue, F. F. Tao, W. Shen, C. J. He, Q. L. Chen, L. J. Wu, Y. M. Zhu, *Chem Phys Lett* **2004**, 385, 477; e) X. L. Zhao, Y. Ando, Y. Liu, M. Jinno, T. Suzuki, *Phys Rev Lett* **2003**, 90; f) W. A. Chalifoux, R. R. Tykwinski, *Nat Chem* **2010**, 2, 967; g) R. R. Tykwinski, W. Chalifoux, S. Eisler, A. Lucotti, M. Tommasini, D. Fazzi, M. Del Zoppo, G. Zerbi, *Pure Appl Chem* **2010**, 82, 891; h) R. R. Tykwinski, *Chem Rec* **2015**, 15, 1060; i) Q. Sun, L. L. Cai, S. Y. Wang, R. Widmer, H. X. Ju, J. F. Zhu, L. Li, Y. B. He, P. Ruffieux, R. Fasel, W. Xu, *J Am Chem Soc* **2016**, 138, 1106.

[7]     P. J. Stang, M. Ladika, *J Am Chem Soc* **1981**, 103, 6437.

[8]     L. Shi, P. Rohringer, K. Suenaga, Y. Niimi, J. Kotakoski, J. C. Meyer, H. Peterlik, M. Wanko, S. Cahangirov, A. Rubio, Z. J. Lapin, L. Novotny, P. Ayala, T. Pichler, *Nat Mater* **2016**, 15, 634.

[9]     a) S. Heeg, L. Shi, L. V. Poulikakos, T. Pichler, L. Novotny, *Nano letters* **2018**, 18, 5426; b) L. Shi, R. Senga, K. Suenaga, H. Kataura, T. Saito, A. P. Paz, A. Rubio, P. Ayala, T. Pichler, *Nano Lett* **2021**, 21, 1096.

[10] a) L. Ravagnan, G. Bongiorno, D. Bandiera, E. Salis, P. Piseri, P. Milani, C. Lenardi, M. Coreno, M. de Simone, K. C. Prince, *Carbon* **2006**, 44, 1518; b) L. Ravagnan, P. Piseri, M. Bruzzi, S. Miglio, G. Bongiorno, A. Baserga, C. S. Casari, A. L. Bassi, C. Lenardi, Y. Yamaguchi, T. Wakabayashi, C. E. Bottani, P. Milani, *Phys Rev Lett* **2007**, 98; c) M. Bogana, L. Ravagnan, C. S. Casari, A. Zivelonghi, A. Baserga, A. L. Bassi, C. E. Bottani, S. Vinati, E. Salis, P. Piseri, E. Barborini, L. Colombo, P. Milani, *New J Phys* **2005**, 7; d) Y. F. Zhang, J. W. Zhao, Y. H. Fang, Y. Liu, X. L. Zhao, *Nanoscale* **2018**, 10, 17824.

[11] a) F. Banhart, A. La Torre, F. Ben Romdhane, O. Cretu, *Eur Phys J-Appl Phys* **2017**, 78; b) N. F. Andrade, T. L. Vasconcelos, C. P. Gouvea, B. S. Archanjo, C. A. Achete, Y. A. Kim, M. Endo, C. Fantini, M. S. Dresselhaus, A. G. Souza, *Carbon* **2015**, 90, 172; c) K. Asaka, S. Toma, Y. Saito, *Sn Appl Sci* **2019**, 1.

[12] A. Milani, M. Tommasini, V. Russo, A. L. Bassi, A. Lucotti, F. Cataldo, C. S. Casari, *Beilstein J Nanotech* **2015**, 6, 480.

[13] a) K. Kang, T. Ozel, D. G. Cahill, M. Shim, *Nano Lett* **2008**, 8, 4642; b) D. H. Song, F. Wang, G. Dukovic, M. Zheng, E. D. Semke, L. E. Brus, T. F. Heinz, *Phys Rev Lett* **2008**, 100; c) H. G. Yan, D. H. Song, K. F. Mak, I. Chatzakis, J. Maultzsch, T. F. Heinz, *Phys Rev B* **2009**, 80; d) J. M. Nesbitt, D. C. Smith, *Nano Lett* **2013**, 13, 416; e) J. Y. Zhu, R. German, B. V. Senkovskiy, D. Haberer, F. R. Fischer, A. Gruneis, P. H. M. van Loosdrecht, *Nanoscale* **2018**, 10, 17975.

[14] R. Hoffmann, *Angew Chem Int Edit* **1987**, 26, 846.

[15] C. S. Casari, M. Tommasini, R. R. Tykwinski, A. Milani, *Nanoscale* **2016**, 8, 4414.

[16] V. W. Brar, G. G. Samsonidze, M. S. Dresselhaus, G. Dresselhaus, R. Saito, A. K. Swan, M. S. Unlu, B. B. Goldberg, A. G. Souza, A. Jorio, *Phys Rev B* **2002**, 66.

[17] a) C. Fantini, E. Cruz, A. Jorio, M. Terrones, H. Terrones, G. Van Lier, J. C. Charlier, M. S. Dresselhaus, R. Saito, Y. A. Kim, T. Hayashi, H. Muramatsu, M. Endo, M. A.

Pimenta, *Phys Rev B* **2006**, 73; b) S. Heeg, L. Shi, T. Pichler, L. Novotny, *Carbon* **2018**, 139, 581.

[18]  L. Shi, P. Rohringer, M. Wanko, A. Rubio, S. Wasserroth, S. Reich, S. Cambre, W. Wenseleers, P. Ayala, T. Pichler, *Phy Rev Mater* **2017**, 1.

[19]  S. Han, C. Boguschewski, Y. Gao, L. Xiao, J. Zhu, P. H. M. van Loosdrecht, *Opt Express* **2019**, 27, 29949.

[20]  a) N. Kamaraju, S. Kumar, Y. A. Kim, T. Hayashi, H. Muramatsu, M. Endo, A. K. Sood, *Appl Phys Lett* **2009**, 95; b) X. L. Zhang, Z. B. Liu, X. Zhao, X. Q. Yan, X. C. Li, J. G. Tian, *Opt Express* **2013**, 21, 25277.

[21]  J. Zhu, R. B. Versteeg, P. Padmanabhan, P. H. M. van Loosdrecht, *Phys Rev B* **2019**, 99.

[22]  P. Rohringer, L. Shi, P. Ayala, T. Pichler, *Adv Funct Mater* **2016**, 26, 4874.

[23]  a) A. Rajan, M. S. Strano, D. A. Heller, T. Hertel, K. Schulten, *J Phys Chem B* **2008**, 112, 6211; b) D. M. Harrah, A. K. Swan, *Acs Nano* **2011**, 5, 647; c) J. H. Yoon, H. Kim, *B Korean Chem Soc* **2017**, 38, 364.

[24]  L. Shi, K. Yanagi, K. C. Cao, U. Kaiser, P. Ayala, T. Pichler, *Acs Nano* **2018**, 12, 8477.

[25]  L. Shi, P. Rohringer, P. Ayala, T. Saito, T. Pichler, *Phys Status Solidi B* **2013**, 250, 2611.

[26]  R. B. Versteeg, J. Zhu, P. Padmanabhan, C. Boguschewski, R. German, M. Goedecke, P. Becker, P. H. M. van Loosdrecht, *Structural dynamics* **2018**, 5, 044301.

# Supporting Information

**(1). Radial breathing modes of single walled carbon nanotubes, double walled carbon nanotubes and double walled carbon nanotubes with confined long linear carbon-chains.**

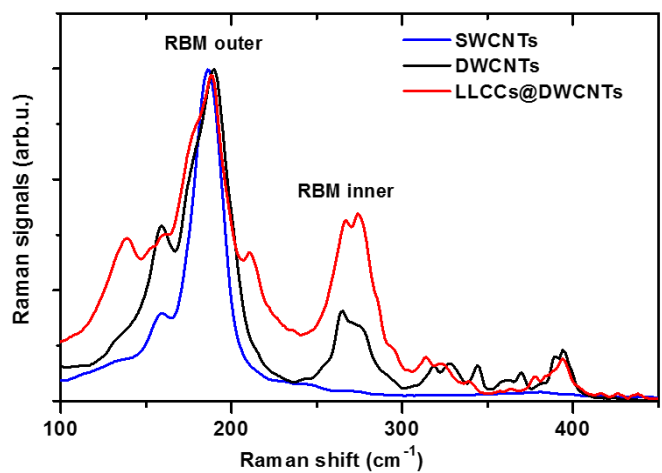

**Figure S1.** Low frequency region radial breathing modes of SWCNTs, DWCNTs and LLCCs@DWCNTs recorded with continuum laser at a wavelength of 532 nm. The vibrational peaks below and above 250 cm$^{-1}$ were ascribed to come from the outer and inner tubes respectively.

**(2). Energy dependent resonance Raman scattering spectra of LLCCs@DWCNTs reordered with femtosecond light source.**

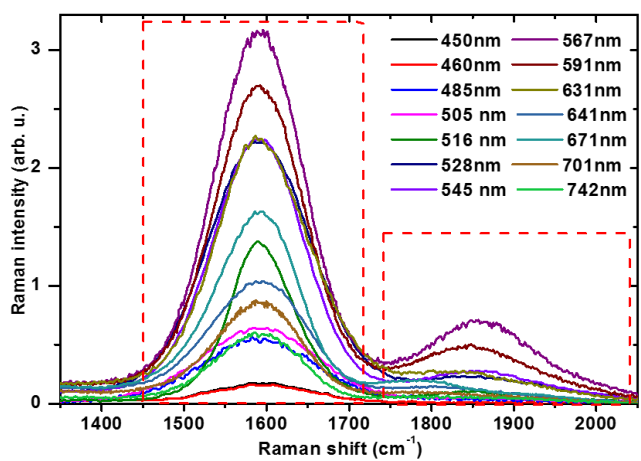

**Figure S2.** Wavelength dependent resonance Raman scattering spectra of the LLCCs@DWCNTs recorded with femtosecond pulse laser. The red dashed line squares indicate the spectral regions for integration of the total

intensity which represent the DWCNTs and CC signals at ~1590 cm$^{-1}$ and ~1850 cm$^{-1}$ respectively in the main manuscript.

## (3). Relaxation dynamics the nanotubes observed from different kinds of samples.

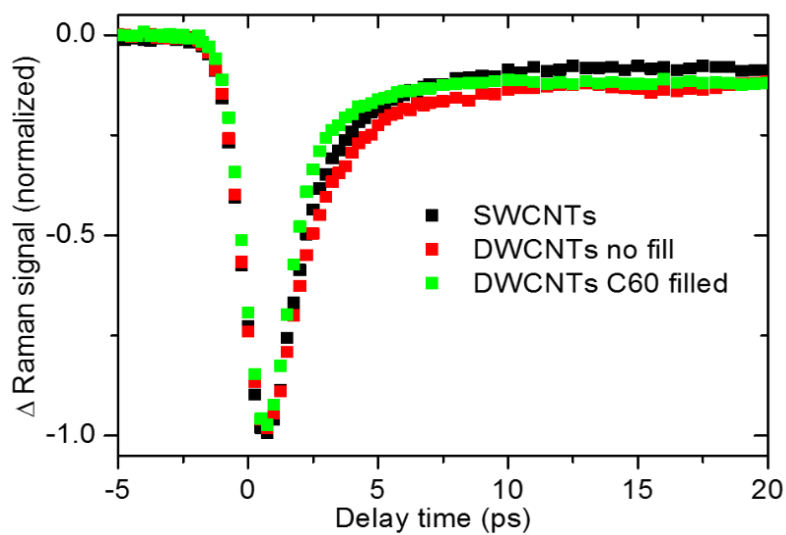

**Figure S3.** Dynamics observed at 1590 cm$^{-1}$ for the eDIPS SWCNTs, DWCNTs obtained by eDIPS SWCNTs annealing above 1800 °C and DWCNTs obtained by annealing eDIPS SWCNTs filled with C60. In this three tube samples, similar relaxation dynamics was observed.

## (4). Log-normal fitting of the decay dynamics of LLCC mode

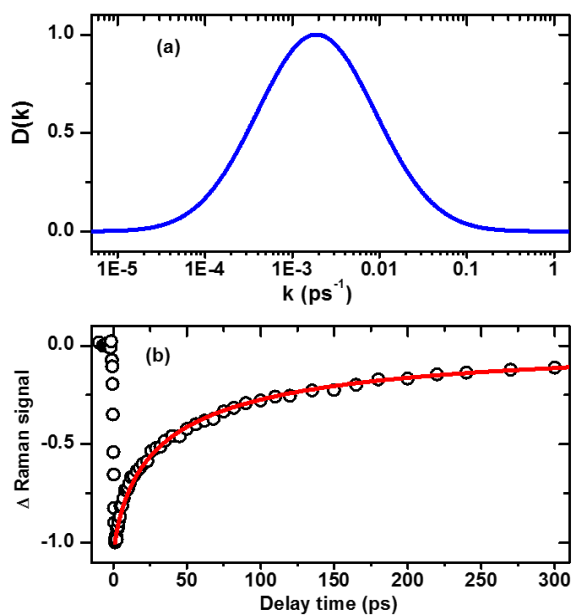

**Figure S4.** Decay rate constant distribution fitting of the exciton dynamics of LLCCs with expression of $\sum_{k=0}^{\infty} D(k)\exp(-kt)$, D(k) is assumed in Log-normal distribution as indicated in (a), and the fitted curve is shown in (b) by the red line.

## (5). Excitation energy dependent dynamics of LLCCs@DWCNTs and global fitting results.

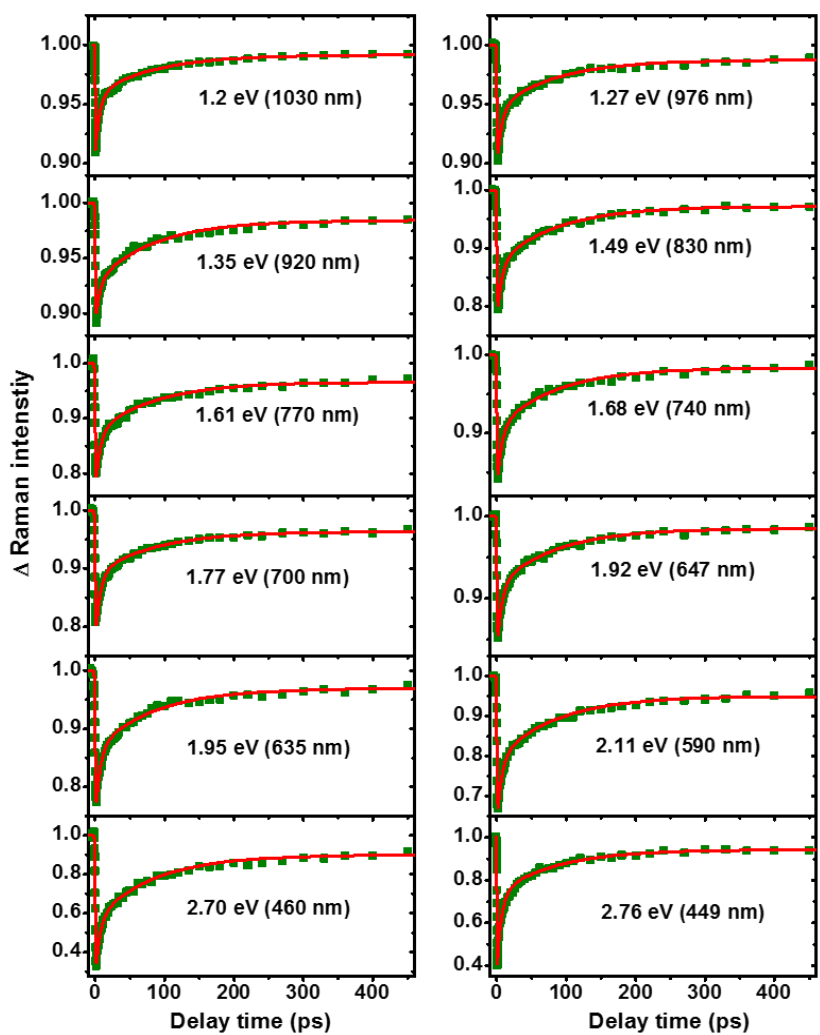

**Figure S5.** Excitation photon energy dependent relaxation dynamics of the confined CC in DWCNTs observed from the vibration peak at ~1850 cm$^{-1}$. Dots are experimental signals and red lines are global fitted ones with multi-exponential decay functions with lifetime constants of 7 ps and 80 ps and non-decay components.

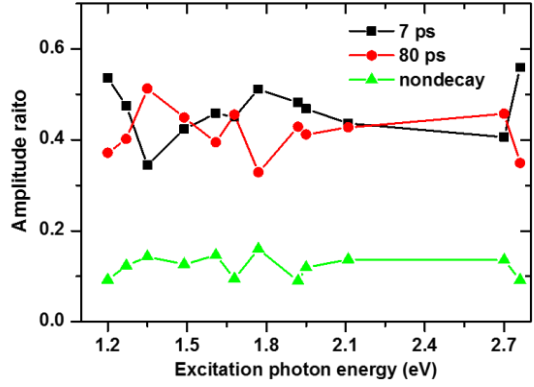

**Figure S6.** Global fitting of the dynamics in figure S4 extracted relative amplitude ratio for each exponential decay components.

## (6). Excitation power dependent of the dynamics of LLCCs@DWCNTs.

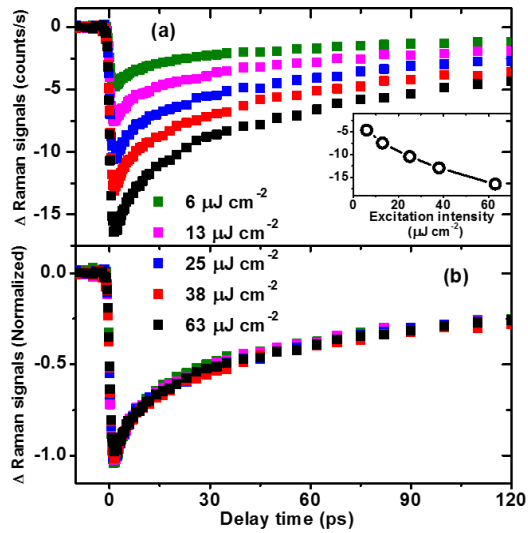

**Figure S7.** Pump intensity dependence of the excited state dynamics of LLCCs@DWCNTs observed through the phonon Raman response at ~1840 cm$^{-1}$. (a) Dynamics at different excitation intensities (unscaled). The inset in (a) shows the pump excitation intensity dependence of the maximum negative amplitude at a delay of around 1 ps. (b) Normalized decay dynamics to show the absence of any non-linear dependence on the pump intensity. Experiments were performed at room temperature with Pump phonon energy is at ~2.1 eV.